\documentstyle[prd,aps,preprint]{revtex}

\begin{document}

\draft

\preprint{Imperial/TP/93-94/34 and hep-ph/9405221}

%\twocolumn[\hsize\textwidth\columnwidth\hsize\csname @twocolumnfalse\endcsname

\title{Non-intercommuting configurations in the collisions of type I
$U(1)$ cosmic strings}
\author{L. M. A. Bettencourt\footnote{e-mail: l.bettencourt@ic.ac.uk} and  T.
W. B. Kibble }
\address{The Blackett Laboratory, Imperial College,
London SW7 2BZ, U.K.}
\date{\today}
\maketitle

\begin{abstract}
It is shown that for  small relative angle and kinetic
energy two type I $U(1)$ strings can form bound states upon collision
instead of the more familiar intercommuting configuration. The
velocity below which this may happen is estimated as function
of the ratio of the coupling constants in the theory, crossing
angle and initial kinetic energy.
\end{abstract}

%\pacs{PACS Numbers : 98.80.Cq}

%\vskip2pc]

\clearpage

Cosmic Strings formed at a grand unified symmetry breaking phase
transition \cite{Kib} are a strong candidate to solve one of the
most outstanding
problems of cosmology namely that of structure seeding \cite{Struc}.
The standard cosmic string scenario assumes that strings after the
phase transition rapidly approach the limit of  zero width and
become henceforth well described in terms of the Nambu-Goto action.
The other necessary ingredient to specify the evolution of the
network is that at each collision strings should exchange ends or in other
words should intercommute. In particular, whenever two segments of the
same long
string intersect a loop will be formed. Loops of string, in turn, are
not globally topologically stable and will shrink by emitting their
energy presumably as gravitational radiation. This process allows for
a faster decrease of the energy density in strings and precludes the
possibility of a string dominated universe.

This whole construction rests crucially on the assumption that the outcome
of a string collision always corresponds to intercommuting.
However, strings are in rigour classical non-trivial solutions of
a variety of field theories displaying spontaneous symmetry breaking
\cite{Kib}.
The preferred model for realising string dynamics, however, is the
abelian Higgs model

\begin{eqnarray}
{\cal L} = -{1 \over 4 } F_{\mu \nu} F^{\mu \nu} + {1 \over 2} \vert
(\partial_\mu + iqA_\mu) \phi \vert ^2 -{\lambda \over 8} (\vert \phi
\vert ^2 - \eta^2)^2.
\end{eqnarray}

\noindent In what follows we will also confine ourselves to
it and its $U(1)$ string solutions \cite{NO}.

The only way to establish intercommuting as the necessary outcome of
the collision between two cosmic strings is to study the dynamics of
the full field theory solutions. This, however, involves solving a set
of non-linear equations for the fields which proves to be a rather
difficult task in most circumstances. One way out is to perform
simulations of collisions by solving the theory numerically. This was
done for several  values of the ratio of the coupling constants in
(1), $\beta = \sqrt{q^2 \over \lambda}$, supporting the
conjecture that global, type II and critically coupled strings (i.e.,
$\beta \leq 1$) intercommute under essentially all circumstances
\cite{She,Shell,Mat}. Similar studies regarding the behaviour of type I
strings are much scarcer and seem to open the possibility for the
existence of more complicated configurations \cite {Lag}.

For a collision between two cosmic strings to be analytically
understood in terms of its
underlying field theory one needs to look, in the first instance,  at
vortex dynamics. Indeed,
both strings, at the collision point, can be projected onto a set of
two orthogonal
planes, thus reducing the problem to the study of vortex dynamics
in two dimensions\footnote{This construction should work particularly
well for type
II strings, when the character of the interactions is dominated by the
gauge field and consequently is sensitive to spatial orientation
\cite{Bett}. The approximation of the dynamics of the whole string to
that of vortices, however, excludes the possible excitation of modes
along the string which can play an important role in dissipative effects.}
\cite{Shell,Cop}. In this picture intercommuting
results from the annihilation of a
vortex-antivortex pair in one of these planes and the scattering at
right angles of the vortex-vortex pair on the other.
The scattering at right angles, in turn, is only
fully understood analytically in the critically coupled regime
\cite{Ruback}.
 This
important result may, although possibly under more severe
restrictions, generalise to non-critically coupled vortices.  As a
practical result the interactions should vanish at zero separation as
is observed numerically \cite{Jac}, at least for $\beta$ not too far
from unity. This conjecture also seems to be
well supported by several other numerical and semi-analytical recent
studies of non-critically coupled vortex scattering dynamics
\cite{Rebbi,Shah}.
It then seems likely that the effect of the interactions between
cosmic strings should be felt at intermediate distances, i.e., of the
order of several scalar field coherence lengths. The character
of these is better known than their short distance counterpart.
In particular,   for $\beta> 1$ (type I) \cite{Bett,Jac}
these interactions are always attractive, regardless of the relative
orientation of a pair of strings. In principle this leads to the
possibility of formation of a bound state (characterised by the sum of
the individual winding numbers of the two original strings) as the
outcome of a sufficiently
favourable collision between two string segments.

In this letter we explore this scenario by investigating  what
configurations could compete
with the usual intercommuting as the outcome of  the collision of
two straight type I strings at an angle and under which
circumstances these can be expected to prevail.

A typical intercommuting event that occurs at the origin at $t=0$ may
be taken to be between two strings whose initial configuration is
given (for $t<0$) by

\begin{eqnarray}
x_1 &=& (t, \xi \sigma, -\eta \sigma, \zeta t) \nonumber \\
x_2 &=& (t, \xi \sigma, \eta \sigma, - \zeta t),
\end{eqnarray}

\noindent where $\xi^2 +\eta^2 +\zeta^2=1$. Note that $\zeta$ is the
velocity of approach. We can write
$ x_j = { 1 \over 2} \left[ a_j(u) + b_j(v) \right] $,
where $u=t+\sigma$ and $v=t-\sigma$.

Then,
\begin{eqnarray}
a_1 &=& u(1,\xi,-\eta,\zeta) \nonumber \\
b_1 &=& v(1,-\xi,\eta,\zeta),
\end{eqnarray}
\noindent and
\begin{eqnarray}
a_2 &=& u(1,\xi,\eta,-\zeta) \nonumber \\
b_2 &=& v(1,-\xi,-\eta,-\zeta).
\end{eqnarray}

If we have the normal type of intercommuting event, the solution for
$t>0$ is easy to describe. For $\vert \sigma \vert >t$, the previous
solutions are still valid, but in the region where $u>0$ and $v>0$ the
partners have swapped over. The solutions are as illustrated in figure
1.
The ``horizontal'' pieces of string, in fig. 1, are described by
\begin{equation}
{1 \over 2} \left[ a_2(u) +b_1(v)\right] = (t,\xi \sigma,\eta t, -
\zeta \sigma),
\end{equation}
\noindent and
\begin{equation}
{1 \over 2} \left[ a_1(u) +b_2(v)\right] = (t,\xi \sigma,-\eta t,
\zeta \sigma).
\end{equation}
Note that the velocity on these sections is in the $y$-direction, not
in the $z$-direction.

Can we then find an alternative solution, based on the assumption that
a segment of $n=2$ string is created at the moment of intersection?
{}From symmetry it is easy to see that if such a segment is created it
should be static and directed along the $x$-axis.
Let us write $\epsilon={2 \mu \over \mu_2}$, where $\mu$ is the
string tension for unit winding number, and $\mu_2$ the corresponding
value for $n=2$. $\epsilon$ is naturally a function of $\beta$ and, in
particular for type I strings, we have $\epsilon(\beta>1) >1$. Then
the solution for the $n=2$ segment should be of the form
\begin{equation}
x_D=(t, \epsilon \sigma,0,0).
\end{equation}
It is reasonable to assume that the ends of this segment must move
with definite velocity. Let us take $x_D$ to be the solution for
$\sigma$ in the range $-\kappa t <\sigma<\kappa t$, for some $\kappa$.
Then the ends will move with speed $\kappa \epsilon$ (obviously $\kappa
\epsilon <1$).
 The overall shape must then be as illustrated in fig. 2, with four links
at the same places as before receding with speed $1$, joined to the
doubled section by new straight segments.
We can then write down the expressions for the string position on each
of the straight segments, since we know the coordinates of the
end-points
\begin{eqnarray}
x_3 &=& {1 \over 1- \kappa} \left((1-\kappa)t, -\epsilon \kappa ( t +
\sigma ) + (\sigma + \kappa t)\xi, -(\sigma + \kappa t) \eta, - (\sigma +\kappa
t)
\zeta) \right) \nonumber \\
x_4 &=& {1 \over 1- \kappa} \left((1-\kappa)t, -\epsilon \kappa ( t +
\sigma ) + ( \sigma + \kappa t)\xi, (\sigma + \kappa t) \eta,  (\sigma +\kappa
t)
\zeta) \right) \nonumber \\
x_5 &=& {1 \over 1- \kappa} \left((1-\kappa)t, \epsilon \kappa ( t - \sigma ) +
( \sigma - \kappa t)\xi, -(\sigma - \kappa t) \eta,  (\sigma - \kappa t)
\zeta) \right) \nonumber \\
x_6 &=& {1 \over 1- \kappa} \left((1-\kappa)t, \epsilon \kappa ( t - \sigma ) +
( \sigma - \kappa t)\xi, (\sigma - \kappa t) \eta, - (\sigma - \kappa t)
\zeta) \right).
\end{eqnarray}
Equivalently we can write,

\begin{eqnarray}
x_3 &=& {1 \over 2} \left( a_3 + b_1 \right), \qquad a_3 = {u \over
1-\kappa} \left( 1-\kappa, -2 \epsilon \kappa + (1+\kappa) \xi,
-(1+\kappa)\eta, - (1+ \kappa )\zeta \right) \nonumber \\
x_4 &=& {1 \over 2} \left( a_4 + b_2 \right), \qquad a_4 = {u \over
1-\kappa} \left( 1-\kappa, -2 \epsilon \kappa + (1+\kappa) \xi,
(1+\kappa)\eta,  (1+ \kappa )\zeta \right) \nonumber \\
x_5 &=& {1 \over 2} \left( a_1 + b_5 \right), \qquad b_5 = {v \over
1-\kappa} \left( 1-\kappa, 2 \epsilon \kappa - (1+\kappa) \xi,
(1+\kappa)\eta, - (1+ \kappa )\zeta \right)\nonumber \\
x_6 &=& {1 \over 2} \left( a_2 + b_6 \right), \qquad b_6 = {v \over
1-\kappa} \left( 1-\kappa, 2 \epsilon \kappa - (1+\kappa) \xi,
-(1+\kappa)\eta,  (1+ \kappa )\zeta \right).
\end{eqnarray}

To satisfy the consistency conditions, we require
\begin{equation}
\left[ 2 \epsilon \kappa - (1+\kappa) \xi \right]^2 + (1+\kappa)^2 \eta
^2 + (1+\kappa)^2 \zeta^2 = (1-\kappa)^2,
\end{equation}
\noindent or equivalently
\begin{equation}
4 \epsilon ^2 \kappa ^2 - 4 \epsilon \kappa(1+\kappa) \xi + (1+ \kappa)^2
= (1-\kappa)^2,
\end{equation}
\noindent which yields
\begin{equation}
\epsilon^2 \kappa - \epsilon (1+\kappa) \xi +1 = 0.
\end{equation}

Thus,  the value of $\kappa$ is given by
\begin{equation}
\epsilon \kappa = { \epsilon \xi -1 \over \epsilon - \xi}.
\end{equation}
Clearly a solution exists, if and only if,
\begin{equation}
\xi > {1 \over \epsilon}.
\end{equation}
This is equivalent to requiring the angle between the strings, $\alpha$, to be
small,
or the velocity of approach, $\zeta$, to be small or both. We have $\xi ^2 +
\eta ^2 =
1-\zeta^2= {1 \over \gamma^2}$ and consequently, (see fig. 1),
\begin{equation}
\xi = {\cos ({\alpha \over 2}) \over \gamma} > {1 \over \epsilon}.
\end{equation}
The limit on the angle is then $\cos({\alpha \over 2}) > {\gamma \over
\epsilon}$.
It is not surprising that a ``zipper'' can exist only for small angles
and/or velocities. In particular, for $\beta= 2$,
$\epsilon^{-1}= 0.921$ \cite{Lag}, and the constraint (15) becomes quite
severe.

%\section{The effect of interactions}

Having shown that the formation of ``zippers'' is possible in the
Nambu-Goto approximation we investigate how the consideration of the
interactions between the two colliding strings alters the
corresponding constraints on the collision parameters (15). In the Nambu-Goto
approximation we have seen that the difference between the tension of
the $n=2$ string when compared to that of the two $n=1$ strings added
together, allowed for the formation of the ``zipper''. In this
approximation such a difference is restricted to the $n=2$
bridge only, disregarding the interactions between the two $n=1$ ``legs'' at
each end of it. By including these interactions we should then be able
not only to render our former construction more realistic but also to relax
the rather tight constraints it imposed.

Down to scalar
coherence length separations it was shown \cite{Bett} that the
interaction energy
between two $n=1$ string  elements at an angle $\alpha$, is given
approximately by
\begin{equation}
E_{\rm int}(d)=  2 \pi \eta^2 \left[  \cos(\alpha) K_0(m_A d)- K_0(m_S d)
\right] ,
\end{equation}
\noindent where $m_A= \eta q$ and $m_S= \eta \sqrt{\lambda}$.

On the other hand each vortex possesses a self-energy $\mu$, which naturally
emerges as the string tension in the Nambu-Goto approximation and can
be seen as its inertial mass for the study of collisions. For general
couplings $\mu$ has to be
computed numerically. A best fit to its dependence on $\beta$ yields
\cite{Hod}
\begin{equation}
\mu= 1.04 \pi {\eta^2 \over (2 \beta^2)^{0.195}},
\end{equation}
\noindent which should be valid to about $5\%$ accuracy, in the range
$$ 0.01 < 2 \beta^2 <100. $$

We can then investigate the existence of bound states by matching the
kinetic energy of  two segments emerging from the collision, in the
center-of-mass frame, to the potential energy of the system at a given
separation.  In so doing we are assuming that only a finite length of
string contributes to the dynamics, at the collision point. We express
this in terms of a simple cutoff, $l_{\rm eff}$, a length along the
string measured from the point of collision and take the
velocity of each string as seen from the center-of-mass frame, $v_{\rm CM}$, to
be constant within such an interval\footnote{We should note
that for points in the vicinity  of the collision point $v_{\rm CM}$,
should be largest and then decreasing to zero smoothly with increasing
distance along the string. Our procedure of introducing a simple
cutoff $l_{\rm eff}$, gives us, therefore, a measure of the total
kinetic energy in the direction of separation, but not of its
distribution along the string.}. This procedure yields
\begin{equation}
\int_0^{l_{\rm eff}} d z  2 \mu  (\gamma_{\rm CM} -1) = 2 \pi \eta^2
\int_0 ^{l_{\rm eff}} dz
\left[ K_0(m_S d(z)) - \cos(\alpha) K_0(m_A d(z)) \right],
\end{equation}

\noindent which, in turn, leads to

\begin{equation}
v_{\rm CM}= \sqrt{a^2 + 2 a  \over (1 + a)^2},
\end{equation}
\noindent where
\begin{equation}
a = {1 \over 1.04 l_{\rm eff} }  (2 \beta^2)^{0.195} \int_0 ^{l_{\rm eff}} dz
\left[ K_0(m_S d(z)) - \cos(\alpha) K_0(m_A d(z)) \right],
\end{equation}
\noindent and
\begin{equation}
d(z)= d_0 + z \tan(\alpha).
\end{equation}
 We see, in particular, that if we take $l_{\rm eff}$ to
infinity the integral in (20) should rapidly converge to a constant while
the inverse proportionality of the velocity relative to $l_{\rm eff}$
ensures that the system will always be free. Conversely, in the limit
of vanishing $l_{\rm eff}$,  the  picture for the collision of two
vortices  emerges, with the consequent drop in dependence on $l_{\rm
eff}$. For $\alpha =0$ the integrand in (20) becomes a constant and
this results trivially.
Physically, $l_{\rm eff}$ should depend on the parameters of the
collision, namely on the angle of approach $\alpha$ and on the
time-scale for the exchange of ends between the two strings.
Determining $l_{\rm eff}$ would then require the detailed knowledge
of the dynamics
 of the
collision, which is unknown. In what follows we will, therefore, leave it
as a free parameter and study the effect of its variation upon $v_{\rm CM}$.

The distance $d_0$ corresponds to the separation at
which the kinetic energy should match the potential energy exactly.
Because we expect expression (16) to break down at distances smaller
than the coherence length of the scalar field \cite{Bett} we take it
to  be $d_0= 2 m_S^{-1}$. This condition allows the system to scatter
before the two segments are brought back together, under the effect of
the interactions. It should also correspond to a value close to the
maximum of the interaction energy, since, for smaller separations, it was
shown numerically \cite{Jac} that it should remain approximately
constant as the separation vanishes, at least for $\beta$ not too far
from unity.

Figure 3 shows a plot of the dependence of the binding velocity,
$v_{\rm CM}$,
 on $\beta$ for
four values of $\alpha$ and $l_{\rm eff}= 20 m_S^{-1}$. We see the strong
dependence on the angle of approach $\alpha$. In particular for
orthogonal strings we expect the contribution from the potential
energy to almost vanish and the system will be free. For intermediate
angles, however, the binding velocity, $v_{\rm CM}$ can be naturally in the
range of
$0.1-0.2 c$. The actual value depends quite sensitively on  $l_{\rm eff}$.
Figure 4 a), b), c) show the dependence of the binding velocity
on $\beta$ and $l_{\rm eff}$ for $\alpha= {\pi \over 6}, {\pi \over 4},
{\pi \over 3}$, respectively.
The dependence of the velocity on the angle of approach is shown in
figure 5, for $\beta=2$ and five values of $l_{\rm eff}=5,20,50,100,200
m_S^{-1}$.
In particular it becomes clear that the velocities only converge to a
negligible value for  $\alpha$  close to $\pi \over 2$.
For intermediate angles the variation is relatively small allowing for a large
range of parameters for which the velocity is relativistic.

Ultimately, one wants to relate the binding velocity (19) with
the approach velocity
of two typical strings in the network.  In particular  the collision
needs to be
reasonably inelastic, since otherwise the system could be trivially
time-reversed, resulting in configurations with  the strings at the
same separation and
speed  before and after the collision. One way of parameterising this
is by the use of an
efficiency  parameter, $\rho$, in the following way.
Just before the two zeros of the scalar field superimpose, for very
small separation, we have argued that the system's energy should be
essentially kinetic. Then the same follows for the configuration just
after the collision. However, the incoming energy will be channeled to
all possible modes of an $n=2$ vortex, among which is the one
corresponding to the separation of the two $n=1$ vortices. We then
take $\rho$ to be the ratio between the initial and final (kinetic)
energies, in the separation mode, before and after the collision.
Consequently, $0 \leq \rho \leq 1$. This, in turn, results in a
relation between
the initial and final velocities, before and after the collision
\begin{equation}
v_f^2={\rho v_i^2 \over 1 - v_i^2(1-\rho)},
\end{equation}
\noindent where, $v_f$ will be taken to be the above $v_{\rm CM}$.
Once thus computed, $v_i$ can  be related to a velocity at a given
separation by taking the effect of the interactions into account, for
a given spatial configuration. The typical velocity quoted for a
network of cosmic strings \cite{Ben,Vach} is about $0.15 c$.

%\section{Conclusions}

Our estimates depend quite
sensitively on two undetermined parameters $l_{\rm eff}$ and $\rho$.
Their values can, in principle, be
found by looking at simulations or by solving the short distance
dynamics of strings more exactly analytically.
If $l_{\rm eff}$ and $\rho$ then turn out to be
large, only a very small fraction of all collisions will have a ``zipper''
as the outcome. Subsequent collisions of these configurations with
other strings may lead to the peeling or ``unzipping'' of these higher
winding number bridges as was shown in simulations by
Laguna and Matzner \cite{Lag}. As such the difference between the evolution of
such a network and those evolving under the effect of intercommuting
alone, should be negligible.
For sufficiently low values of $l_{\rm eff}$
and $\rho$, however, we have shown that  a large fraction of
all collisions of strings in the network would lead to local higher
winding number bridges that could grow as ``zippers''.
The overall evolution of such a network can then be considerably
different. In principle loop formation under these circumstances could
still occur, at the same rate, but such loops would remain connected
to their mother string. This should have an effect on their subsequent
evolution.

A network of type I strings can then be richer in string
configurations than one which evolves simply by intercommuting at every
collision. This should have consequences for the evolution of the
characteristic lengths in a string network, which, in turn, could reflect
in the way strings seed energy distribution anisotropies, hopefully
with potentially observable consequences both in  structure in the
universe  and anisotropies in the Cosmic Microwave Background Radiation.

\section*{Acknowledgements}

We would like to thank  Andy Albrecht, Richard Matzner
and Ray Rivers for extremely useful discussions. L.M.A.B.'s
research was supported by J.N.I.C.T. under contract BD 2243/RM/92.

\clearpage
{\large \bf{Figure Captions}}

\vspace {.5in}

{\bf Figure 1} : The intercommuting configuration.

\vspace {.1in}

{\bf Figure 2} : The ``zipper''.

\vspace {.1in}

{\bf Figure 3} : Escape velocity dependence on the ratio of the
couplings, $\beta$,
for $l_{\rm eff}=20 m_S^{-1}$ and
relative angle $\alpha=0^\circ,30^\circ,45^\circ,60^\circ$.

\vspace {.1in}

{\bf Figure 4} : Escape velocity dependence on $\beta$
for $ l_{\rm eff}=5,20,50,100 m_S^{-1} $ and
relative angle

a) $\alpha=30^\circ$

b) $\alpha=45^\circ$

c) $\alpha=60^\circ$

\vspace {.1in}

{\bf Figure 5} : Escape velocity dependence on the angle of approach
$\alpha$  for $l_{\rm eff}=5, 20, 50, 100, 200 m_S^{-1}$ and $\beta=2$.

%\baselineskip=18pt
%%%%%%baselineskip=24pt

\end{document}